\documentclass[aip,reprint]{revtex4-1}

\usepackage{lipsum}
\usepackage{amsmath}
\usepackage{amssymb}
\usepackage{graphicx}
\usepackage{gensymb}
\usepackage{tabularx}
\usepackage{float}
\usepackage{xcolor}
\usepackage{mathtools}
\usepackage{dsfont}
\usepackage[capitalise]{cleveref}
\DeclareUnicodeCharacter{2009}{ }
\DeclareUnicodeCharacter{2192}{$\rightarrow$}
\DeclareUnicodeCharacter{2032}{'}
\DeclareUnicodeCharacter{03B1}{$\alpha$}
\DeclareUnicodeCharacter{03B3}{$\gamma$}
\restylefloat{table}
\graphicspath{ {Images/} }

\begin{document}
	\title{Constraints on Ion Velocity Distributions from Fusion Product Spectroscopy}
	\author{A. J. Crilly}
	\affiliation{Centre for Inertial Fusion Studies, The Blackett Laboratory, Imperial College, London SW7 2AZ, United Kingdom}
	\author{B. D. Appelbe}
	\affiliation{Centre for Inertial Fusion Studies, The Blackett Laboratory, Imperial College, London SW7 2AZ, United Kingdom}
	\author{O. M. Mannion}
	\affiliation{Sandia National Laboratories, Albuquerque, New Mexico 87185, USA}
	\author{W. Taitano}
	\affiliation{Los Alamos National Laboratory, Los Alamos, New Mexico 87545, USA}
	\author{E. P. Hartouni}
	\affiliation{Lawrence Livermore National Laboratory, Livermore, California 94550, USA}
	\author{A. S. Moore}
	\affiliation{Lawrence Livermore National Laboratory, Livermore, California 94550, USA}
	\author{M. Gatu-Johnson}
	\affiliation{Massachusetts Institute of Technology, Plasma Science and Fusion Center, Cambridge, MA 02139, USA}
	\author{J. P. Chittenden}
	\affiliation{Centre for Inertial Fusion Studies, The Blackett Laboratory, Imperial College, London SW7 2AZ, United Kingdom}
	\begin{abstract}
	Recent inertial confinement fusion experiments have shown primary fusion spectral moments which are incompatible with a Maxwellian velocity distribution description. These results show that an ion kinetic description of the reacting ions is necessary. We develop a theoretical classification of non-Maxwellian ion velocity distributions using the spectral moments. At the mesoscopic level, a monoenergetic decomposition of the velocity distribution reveals there are constraints on the space of spectral moments accessible by isotropic distributions. General expressions for the directionally dependent spectral moments of anisotropic distributions are derived. At the macroscopic  level, a distribution of fluid element velocities modifies the spectral moments in a constrained manner. Experimental observations can be compared to these constraints to identify the character and isotropy of the underlying reactant ion velocity distribution and determine if the plasma is hydrodynamic or kinetic.
	\end{abstract}
	
	\maketitle

	\section{Introduction}
	In thermonuclear fusion, the large Coulomb barrier leads to the small population of high energy ions being the dominant reactants. The reaction rates and product energies are therefore dependent on form of the reactants' velocity distribution, especially for the high energy tails. The strong converging shocks in Inertial Confinement Fusion (ICF) experiments are expected to give rise to ion kinetic effects \cite{Rinderknecht2015} and suprathermal fusion products upscatter reactants, populating the tails of the velocity distribution \cite{Hayes2015}. 
	
	The products of fusion reactions are used to diagnose the conditions in ICF experiments. This is often done with the neutrons from the D(T,n)$\alpha$ and D(D,n)$^3$He reactions. Fusion product spectroscopy is a particularly powerful technique as it is uniquely sensitive to the ion velocity distribution. However in current analysis\cite{Hatarik2015}, inference of thermodynamic ion temperatures requires an assumption of equilibrium reactant distributions, such as the formulae given by Brysk\cite{Brysk1973}. The effect of non-equilibrium distributions needs to be further investigated to help understand experiments where ion kinetic effects play a large role.
	
	In this work, we will consider the fusion reactions between two reactant distributions which are not in equilibrium i.e. the distributions are not Maxwellians at the same temperature. The case studies given in this analysis will be focussed towards neutron spectroscopic signals. However, the results are general and can be applied to any 2-to-2 body fusion reaction products e.g. the protons from D($^3$He,p)$\alpha$. We will consider both isotropic and anisotropic velocity distributions to provide the most general theory possible -- isotropy here means the distribution function has full rotational symmetry in velocity space and is therefore a function of particle speeds only.

	\section{Theory of Spectral Moments}

	Given the reactant distributions it is possible to numerically calculate the fusion product spectra\cite{Goncharov2015,Appelbe2011,Higginson2019,Appelbe2021}. However, this can be a computationally intensive task for the full 6D integral and does not lend itself well to understanding experimentally measured spectra. Instead, it is preferable to evaluate the moments of the fusion product spectra, without specifying its functional form. Simple fitting models can incorporate these moment calculations and are well suited to experimental analysis. We will therefore build upon the analysis of Ballabio \textit{et al.} \cite{Ballabio1998} which calculates the product spectra moments for single ion temperature reactants, with relativistic corrections. We will restate the relevant background theory and results here before considering non-equilibrium cases.
	
	For the 2-to-2 reaction $1+2\rightarrow3+4$, the outgoing energy of product particle 3 is given by:
	\begin{align}\label{eqn:classicalkinematicE3}
	E_3 &= \frac{1}{2}m_3v_{cm}^2+\frac{m_4}{m_3+m_4}\left(Q+K\right)+ \\ &v_{cm}\left(\frac{2m_3m_4}{m_3+m_4}\left(Q+K\right)\right)^{\frac{1}{2}}\mu' \nonumber \, \\
	K &\equiv \frac{1}{2}\frac{m_1m_2}{m_1+m_2}  \left| \vec{v}_1-\vec{v}_2 \right|^2
	\end{align}
	where $\vec{v}_i$ and $m_i$ are the velocity and mass of the $i$-th species, $\vec{v}_{cm}$ is the centre of mass (CoM) velocity, $K$ is the relative kinetic energy, $Q$ is the energy released in the reaction and $\mu'$ is the cosine of the angle between the CoM velocity and CoM product velocity. The last term of this expression will vanish when averaging over isotropic distributions \cite{Ballabio1998}. For anisotropic distributions, this term cannot be neglected. In this introductory section we will focus on the well-established isotropic theory. The spectral moment analysis has been extended to consider anisotropic distributions in \cref{section:aniso}.

	By averaging over the reaction rate, one can then derive expressions for the mean and variance of the outgoing particle.
	\begin{align}
	\langle E_3 \rangle &= \alpha_0 + \alpha_K \langle K \rangle + \alpha_V \langle v_{cm}^2 \rangle + ... \ , \\
	\langle E_3^2 \rangle - \langle E_3 \rangle^2 &= \beta_0 \langle v_{cm}^2 \rangle + ...  \ ,
	\end{align}
	where $\langle \rangle$ indicates a reaction rate average:
	\begin{align}\label{eqn:reactionrate}
		\langle x \rangle &= \frac{1}{\langle \sigma v \rangle}\int d^3v_1 f_1(v_1) \int d^3v_2 f_2(v_2) \cdot x \sigma \left| \vec{v}_1-\vec{v}_2 \right| \ , \\
		\langle \sigma v \rangle &= \int d^3v_1 f_1(v_1) \int d^3v_2 f_2(v_2) \sigma \left| \vec{v}_1-\vec{v}_2 \right| \label{eqn:reactivity}
	\end{align}
	where $\sigma$ and $\langle \sigma v \rangle$ the fusion cross section and reactivity respectively. The distribution function of the $i$-th species is denoted $f_i$. The coefficients $\alpha_n$ and $\beta_n$ are functions of the particle masses only and have been calculated by Ballabio \textit{et al.} \cite{Ballabio1998}.
	
	Traditionally, when analysing the fusion product spectra, measurements are made of the shift and variance of the spectral peak. Only the dominant terms for these properties are kept in the following:
	\begin{align}
	\Delta E &= \alpha_K \langle K \rangle + \alpha_V \langle v_{cm}^2 \rangle \ , \label{eqn:specmeanshiftdef}\\
	\sigma^2 &= \beta_0 \langle v_{cm}^2 \rangle \ . \label{eqn:specwidthdef}
	\end{align}
	Therefore only the averages of $K$ and $v_{cm}^2$ are needed to calculate the cumulants of the primary spectrum to first order. It has been shown for a stationary Maxwellian that the neutron spectral variance (or width) is a measure of thermodynamic temperature \cite{Brysk1973}. We will use this fact to define a spectral temperature as follows:
	\begin{equation}
	T_s = \frac{(m_1+m_2)}{3\beta_0} \sigma^2 = \frac{(m_1+m_2)}{3} \langle v_{cm}^2 \rangle \ . \label{eqn:spectempdef}
	\end{equation}
	In the case of a stationary Maxwellian it will, by definition, coincide with the thermodynamic temperature but this is a special case. Another unique property of single temperature Maxwellians is that they exhibit no correlation between $K$ and $v_{cm}^2$. For more general distributions a correlation will exist, introducing a cross section dependence to $\langle v_{cm}^2 \rangle$.
	
	These changes to spectral peak mean and variance can be understood through the physical mechanisms of energy conservation and Doppler broadening respectively. First, the mean shift occurs as the energy from the reactants must be passed onto the products. The energy in the reactants is split between the relative kinetic energy, $K$, and the energy associated with the motion of the CoM. Fusion cross sections are strong increasing functions of $K$ due to the Coloumb barrier penetrability, which exponentially suppresses the cross section at low energies. Therefore, reactions will preferentially occur at large $K$.
	
	\begin{figure}[t]
		\centering
		\includegraphics*[width=0.86\columnwidth]{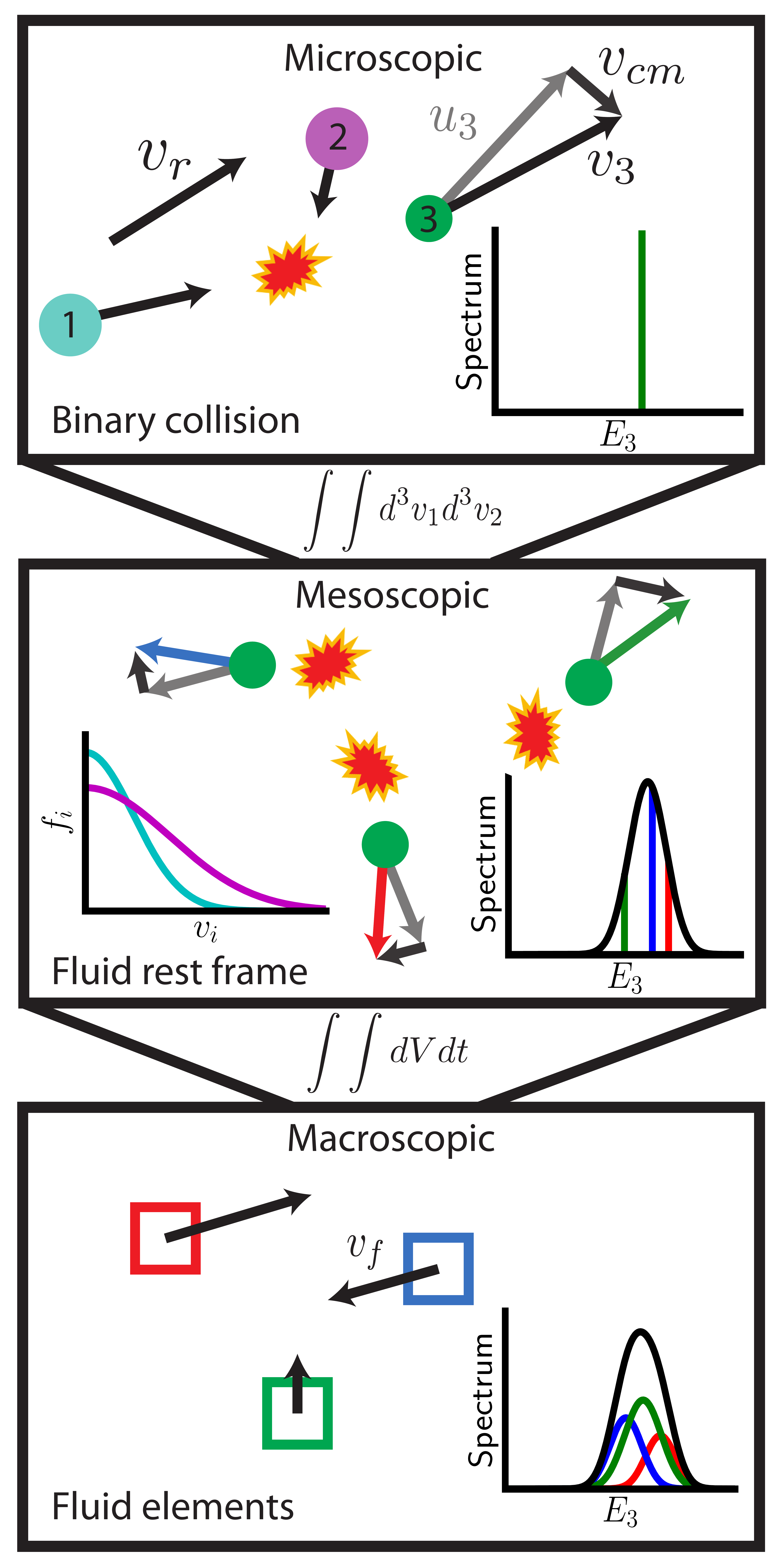}
		\caption{Diagram showing how fusion product spectra are determined at increasingly larger spatial scales. At the microscopic scale, we show a fusion reaction between particles of species 1 and 2 (in cyan and purple) emitting a product of species 3 (in green) at an energy determined by the relative and centre of mass velocities and reaction Q. At the mesoscopic scale, species 1 and 2 have velocity distributions (in cyan and magnetic) producing a spectrum of fusion products. Three example reactions are shown with final product velocities shown in green, red and blue. At the macroscopic scale, we have a collection of fluid elements each emitting their own local spectra. The fluid velocities of these fluid elements introduce Doppler shifts to the emitted spectra. The total spectrum is the sum of all the local fluid element spectra.}
		\label{fig:FrameDiagram}
	\end{figure}
	
	Purely anti-parallel, matched speed, collisions ($\vec{v}_1 = -\vec{v}_2$) will have the highest reaction probability and a low CoM velocity, however, the occupied phase space volume available for these reactions is typically low. This competition between maximising the reaction cross section and the number of available particle pairs leads to the ``average'' reaction occurring between particles of different momenta leading to a non-zero $v_{cm}$. This CoM velocity causes each emitted fusion product to pick up an individual Doppler shift from the CoM motion. Due to the range of CoM velocities, the summed effect creates Doppler broadening of the spectral peak. This Doppler broadening has traditionally be attributed to thermal temperature \cite{Brysk1973,Ballabio1998} and fluid velocity variance \cite{Appelbe2014,Munro2016}, however these are just two specific manifestations of how a range of CoM velocities is produced. In summary, the product spectra depends on the velocity distribution of the reactants which have sufficient energy to react. The mean of the product energies is sensitive to the average relative and CoM kinetic energy in reacting particle pairs while the variance is only sensitive to the range of CoM velocities. For isotropic distributions, these relationships can be calculated using \cref{eqn:specmeanshiftdef,eqn:specwidthdef}.
	
	So far, we have shown that at the microscopic scale of a binary collision the reaction kinematics determines the product energies. At the mesoscopic scale we consider a local, spatial point where there is a velocity distribution of reactants. The distribution of velocities combined with the microscopic reaction kinematics gives rise to a product energy spectrum. The product spectral moments at the mesoscopic scale depend on reaction rate averages of the microscopic scale kinematics. At the macroscopic scale of a laboratory experiment we expect the reactant velocity distributions to vary spatially and temporally. Each discrete spatial location, or fluid element, produces its own product energy spectrum. Reactions occur locally so each fluid element's emitted spectrum depends only on its own local reactant velocity distribution. The fluid elements may also have fluid velocities introducing additional Doppler shifts. The total emitted spectrum is the summation of all fluid elements' local spectra. Thus, moving between scales introduces an additional level of averaging. This process is summarised in \cref{fig:FrameDiagram}, no description of ICF fusion product spectra is complete unless all scales are accurately captured.

	\section{Interpretation of Spectral Moments}
	
	\subsection{Mesoscopic scale}
\begin{figure}[htp]
		\centering
		\includegraphics*[width=0.99\columnwidth]{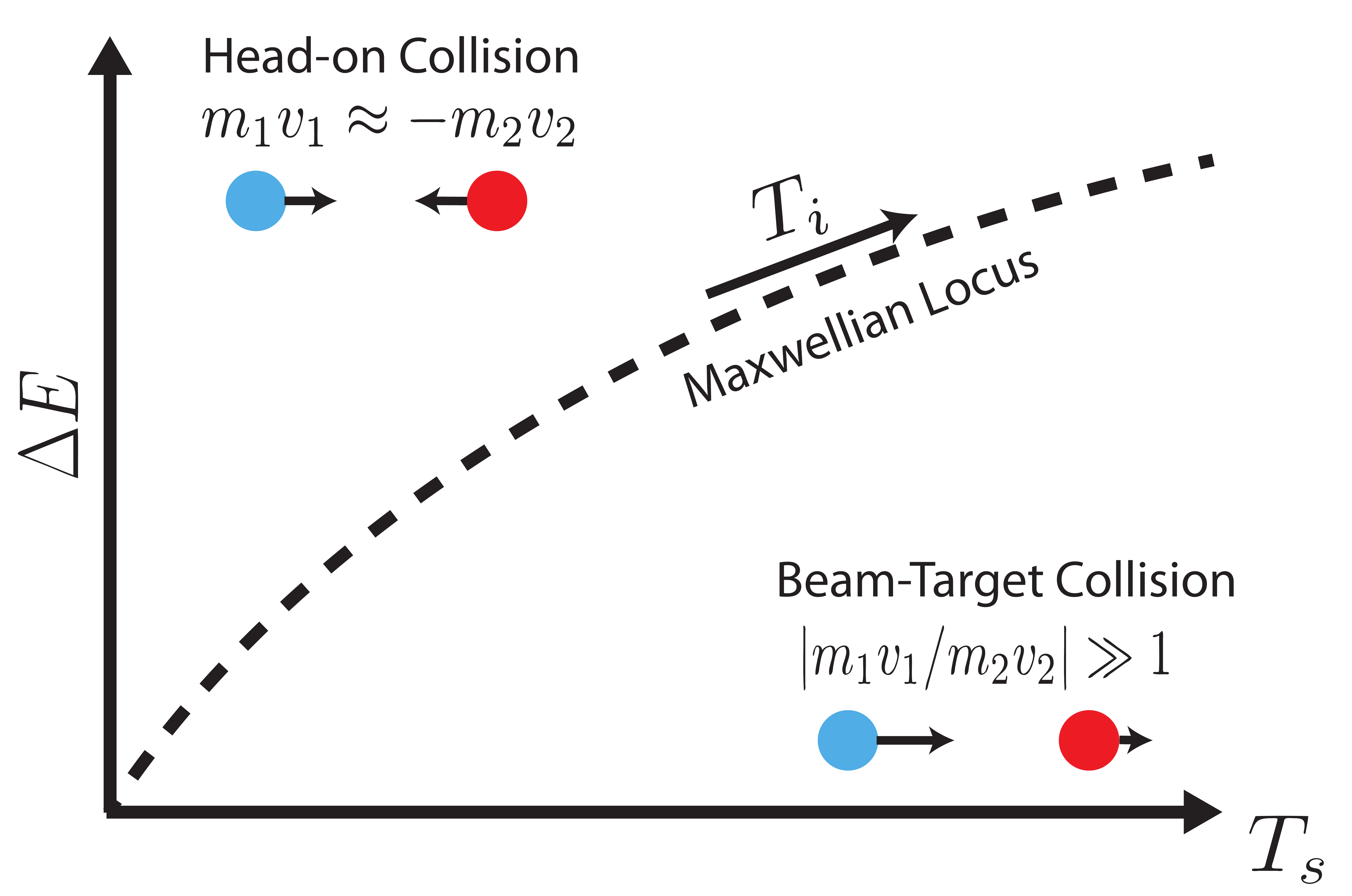}
		\caption{Diagram of the ($\Delta E$, $T_s$) spectral moment space. Collisions of high $K$ and low $v_{cm}^2$ produce spectra with moments in the upper left. Collisions of low $K$ and high $v_{cm}^2$ lie in the lower right - assuming that the distribution of $\vec{v}_{cm}$ is isotropic. Also shown is the Maxwellian locus which gives a 1-to-1 relationship between $\Delta E$ and $T_s$. Changing the thermodynamic temperature simply moves one along this locus and the rest of the space remains inaccessible.}
		\label{fig:SpecMomDiagram}
	\end{figure}

	The spectral moments of fusion product spectra are directly related to the reactant kinematics. A large range in CoM velocities leads to Doppler broadening and an increased spectral temperature. High relative velocities, or relative kinetic energies $\langle K \rangle$, cause an upshift in the spectral mean. Therefore, we can learn about the reactant distributions by observing where their product spectra lie in ($\Delta E,T_s$) space. When considering a single reaction, momenta-matched collisions have zero CoM velocity at finite $K$. Thus, reactant distributions which maximise the number of such collisions will exhibit a large $\Delta E$ and low $T_s$. The opposite is then true for distributions that have large momenta disparity in their collisions, creating a beam-target like reaction. We can therefore use the coordinates in ($\Delta E,T_s$) space to comment on the character of the collisions. \Cref{fig:SpecMomDiagram} shows these observations graphically with the Maxwellian locus between these extrema.

	\subsubsection{Isotropic Distributions}
	
	Isotropic distributions are those which are functions of energy only and therefore are fully symmetric in velocity space. To expand our understanding of spectral moments to all isotropic ion velocity distributions we will consider isotropic monoenergetic distributions. These distributions are shells in velocity space and therefore any general isotropic distribution can be constructed from a superposition of these shells. By varying the relative speed of the reactants we can access the limiting cases of momenta-matching and beam-target discussed above. It is important to note that, as these distributions are isotropic, one cannot ensure all collisions have zero CoM velocity as all collision angles are equally likely. Thus, these monoenergetic distributions produce non-monoenergetic spectra with Doppler broadening produced by the range of collision angles.

\begin{figure}[htp]
		\centering
		\includegraphics*[width=0.99\columnwidth]{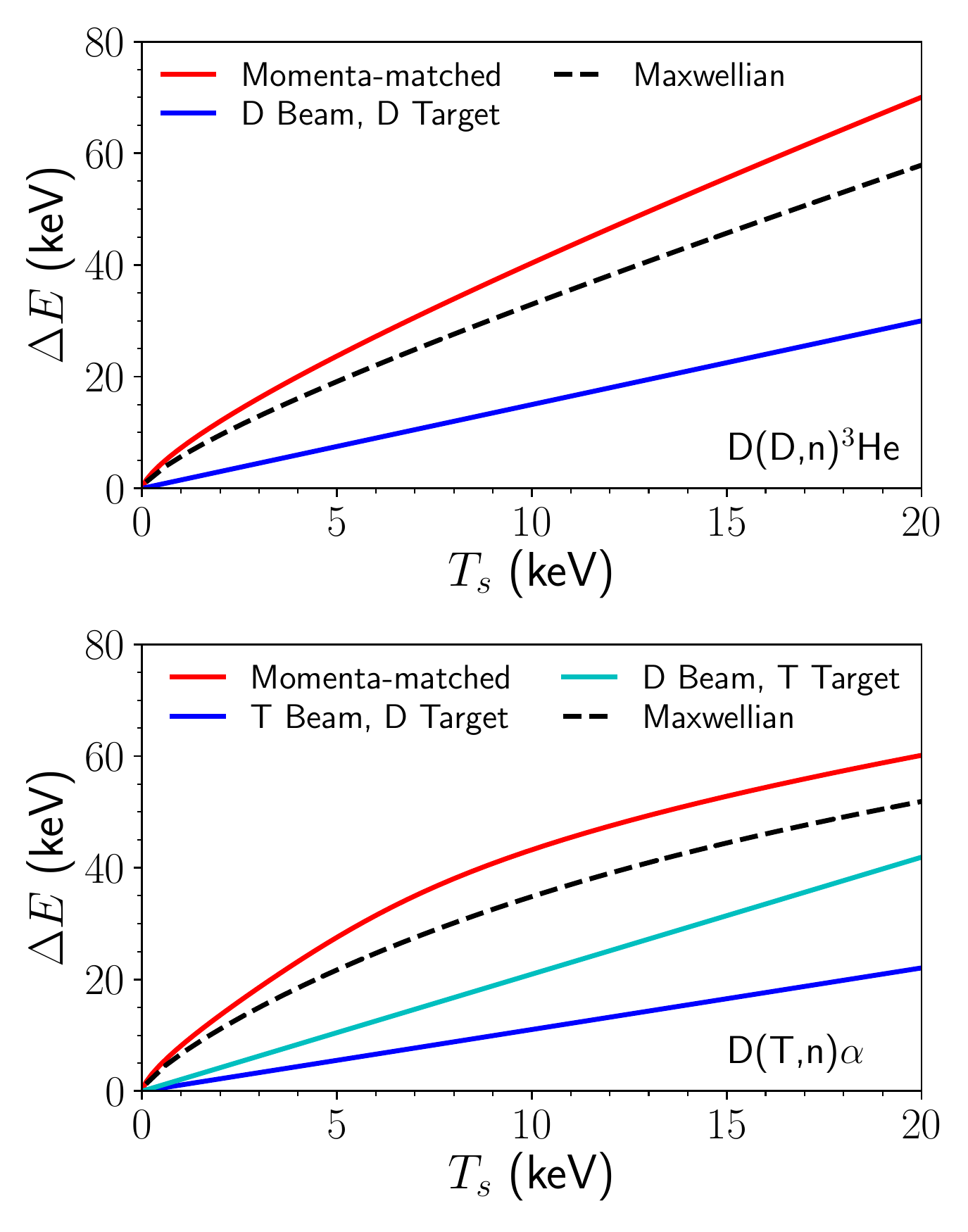}
		\caption{The spectral moments for (top) D(D,n)$^3$He and (bottom) D(T,n)$\alpha$ for limiting cases of monoenergetic distributions. The red lines show the momenta-matched ($m_1|\vec{v}_1| = m_2|\vec{v}_2|$) cases, the blue and cyan lines the beam-target ($v_1 = 0$ or $v_2 = 0$) cases. The Maxwellian locus from Ballabio\cite{Ballabio1998} is shown with a dashed line on both plots.}
		\label{fig:MonoEnergeticDDDT}
	\end{figure}

	\Cref{fig:MonoEnergeticDDDT} shows the (neutronic) spectral moments for the D(D,n)$^3$He and D(T,n)$\alpha$ reactions for the limiting cases of monoenergetic distributions. Full derivation of the spectral moments of monoenergetic distributions is given in \cref{appendix:monodist}. As shown in \cref{fig:MonoEnergeticDDDT}, for the reactions and energy ranges we have considered, these limiting cases create a convex area in ($\Delta E,T_s$) space. This convexity originates from the shape of the fusion cross section. Given these convex boundaries, any point lying on a line connecting two points within these limits will also lie within the limits, this is a restatement of Jensen's inequality\cite{Jensen07}. It is also possible to decompose any other isotropic distribution into an infinite series of monoenergetic shells. The reactivity weighted moments for all the shells will recover the moments of the original distribution. This, combined with the convexity of the monoenergetic shell limits, shows that all isotropic distributions will have spectral moments which lie between the monoenergetic distribution limits, see \cref{fig:MonoEnergeticDDDT}. 

	\begin{figure}[htp]
		\centering
		\includegraphics*[width=0.99\columnwidth]{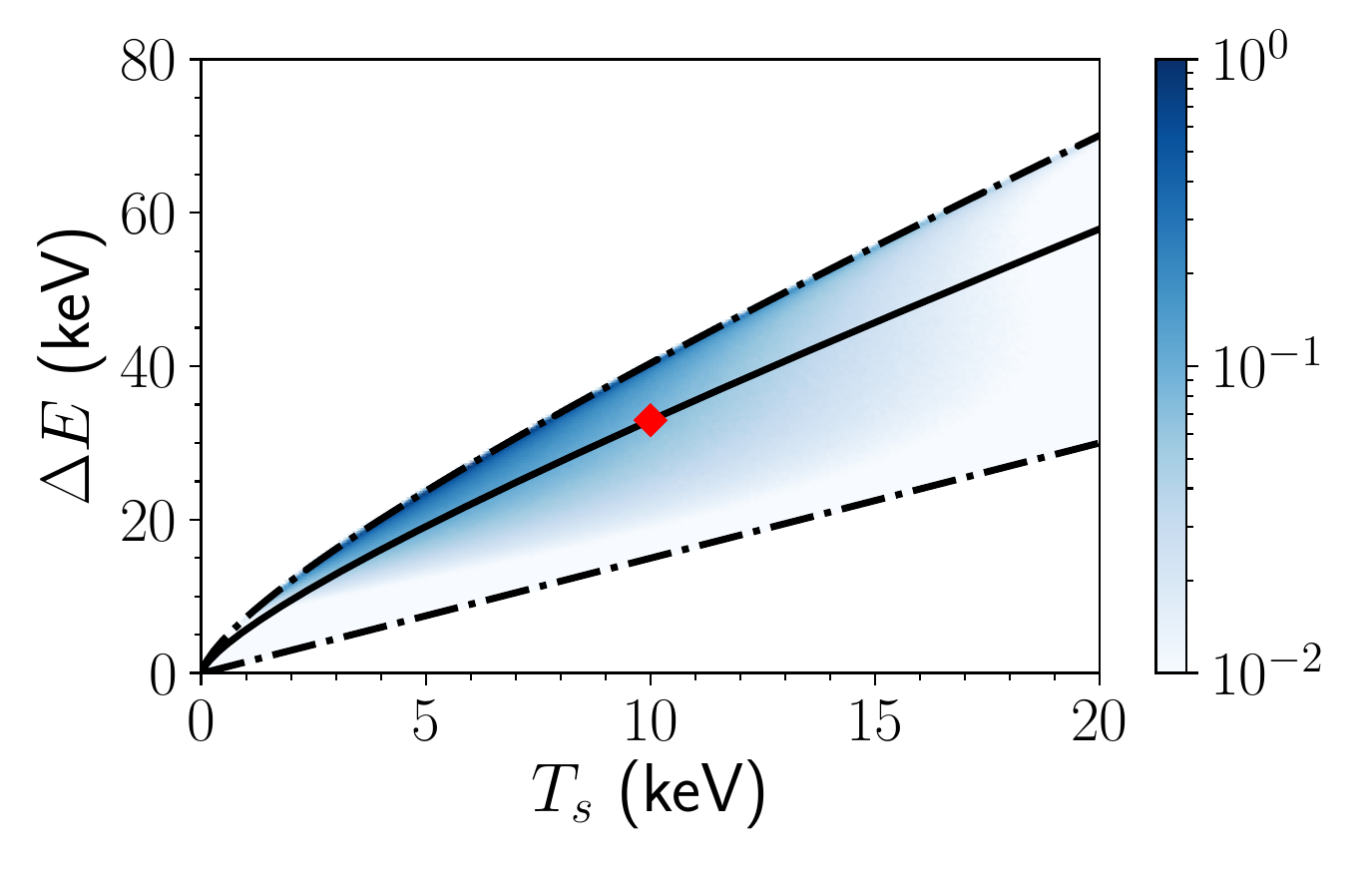}
		\caption{A colour plot showing the monoenergetic shell spectral moment decomposition of a 10 keV D Maxwellian undergoing D(D,n)$^3$He reactions. The colour scale denotes the value of the reactivity-weighted probability distribution function -- this was obtained via a Monte Carlo sampling of the Maxwellian and has been normalised to its peak value. The red diamond denotes the centroid of the PDF. The dot-dashed curves show the limiting cases for monoenergetic shells and the solid line shows the Maxwellian locus. }
		\label{fig:monoenergetic_decomp}
	\end{figure}

	To illustrate this result, we decompose a Maxwellian into monoenergetic shells and calculate the spectral moments for these shells. These necessarily lie within the limits calculated above. We then form a probability distribution function (PDF) of ($\Delta E,T_s$) weighted by the reactivity. The centroid of the 2D PDF returns the spectral moments of the original Maxwellian distribution. The results of this analysis are shown in \cref{fig:monoenergetic_decomp}. It is seen that a large fraction of the reactions occur close to the momentum-matched condition. This leads the Maxwellian locus to lie close to the isotropic distribution upper limit.
	
	We cannot expect spectral moments to be unique properties of given reactant distributions as the 6D velocity space is reduced down to a single coordinate in ($\Delta E,T_s$). Indeed, we have found all isotropic distributions only occupy a fraction of ($\Delta E,T_s$) space and degeneracy is exhibited in \cref{fig:monoenergetic_decomp}. We can however determine distribution function properties such as isotropy and the proportion of momenta-matched collisions from spectral moments.
	
	\subsubsection{Anisotropic Distributions}\label{section:aniso}
	Anisotropic distributions will be able to access a wider range of ($\Delta E,T_s$) space. A trivial example of this is two linear beams of reactants colliding head-on which will exhibit no Doppler broadening but will have a mean shift dependent on the relative kinetic energy. However, the spectral moment analysis becomes more complex when relaxing the isotropy condition as the angular term in the product energy (c.f. \cref{eqn:classicalkinematicE3}) will not necessarily average to zero. Therefore, the fusion product spectrum itself can become anisotropic. While calculating the spectrum from anisotropic distributions is computationally tractable\cite{Goncharov2015, Appelbe2021}, we will instead consider the spectral moment approach for general reactant velocity distributions.
	
	In a similar fashion to Munro's work\cite{Munro2016} on Maxwellian spectra, one can derive expressions for the spectral moments which depend on the emission direction. We will expand on this technique by considering general velocity distributions. The expansion parameter will be the ratio of reactant velocities and the fusion product velocity which we expect to be small as the thermal scale is order keV and nuclear scale is order MeV. The full derivation is given in \cref{appendix:aniso}, we will quote the main results here for the directionally dependent reactivity,
	\begin{subequations}
		\begin{equation}
		\{ \sigma v \} = \langle \sigma v \rangle \left(1+\frac{2}{v_0} \Omega^T \langle \vec{v}_{cm} \rangle + \mathcal{O}\left(\frac{1}{v_0^2}\right)\right) \ , \label{eqn:anisoreactivity}
		\end{equation}
		mean shift,
		\begin{align}
		\{\Delta E\} &= \alpha_K \langle K \rangle + \alpha_V \langle v_{cm}^2 \rangle  \\
		&+ m_3 v_0 \hat{\Omega}^T \langle \vec{v}_{cm} \rangle \nonumber \\
		&- m_3 \hat{\Omega}^T \mathbf{M}(\langle \vec{v}_{cm} \rangle) \hat{\Omega} - \frac{1}{3}m_3|\langle \vec{v}_{cm} \rangle|^2 \nonumber \\
		&+ 3 m_3 \hat{\Omega}^T \langle \mathbf{M}(\vec{v}_{cm}) \rangle \hat{\Omega} + \mathcal{O}\left(\frac{1}{v_0}\right) \ , \nonumber
		\end{align}
		and spectral temperature:
		\begin{align}
		\{T_s\} &= \frac{m_1+m_2}{3} \langle v_{cm}^2 \rangle \\
		&- (m_1+m_2) \hat{\Omega}^T \mathbf{M}(\langle \vec{v}_{cm} \rangle) \hat{\Omega} - \frac{m_1+m_2}{3}|\langle \vec{v}_{cm} \rangle|^2 \nonumber \\
		&+ (m_1+m_2)\hat{\Omega}^T \langle  \mathbf{M}(\vec{v}_{cm}) \rangle \hat{\Omega} + \mathcal{O}\left(\frac{1}{v_0}\right) \nonumber \ ,
		\end{align}
		where we define the $M(\vec{x})$ operator which constructs a rank 2 symmetric traceless tensor from input vector $\vec{x}$:
		\begin{equation}
		\mathbf{M}(\vec{x})  = \vec{x} \otimes \vec{x} - \frac{|\vec{x}|^2}{3} \mathds{1} \ ,
		\end{equation}
	\end{subequations}
	and $\{\}$ denotes a directionally dependent spectral moment, $\hat{\Omega}$ is the emission direction and $v_0$ is the fusion product velocity at $K = v_{cm} = 0$. The tensorial expansion used above is equivalent to a spherical harmonic expansion \cite{Ehrentraut1998} where the angular mode of the terms can be determined by the order in emission direction i.e. terms with no $\hat{\Omega}$ dependence are isotropic, linear in $\hat{\Omega}$ are mode 1, quadratic in $\hat{\Omega}$ are mode 2, etc.. In this analysis we have neglected the anisotropic parts of the differential reaction cross section -- this is a good assumption for the ion energy range of interest for current ICF experiments. Thus, the anisotropy created in the spectrum is purely a result of the Doppler shifts from CoM velocities.
	
	A few key observations can be made from these directionally dependent moment expressions. Firstly, the 4-$\pi$ averages of the reactivity, mean shift and spectral temperature return the same expressions as for isotropic distributions, i.e. \Cref{eqn:reactivity,eqn:specmeanshiftdef,eqn:spectempdef}. Note that for the mean shift and spectral temperature the 4-$\pi$ average requires weighting by the directional yield/reactivity, see \cref{appendix:aniso} for details. Secondly, if the reacting ions have a net CoM velocity, $\langle \vec{v}_{cm} \rangle \ne 0$, then spherical harmonic modes 1 and 2 terms are introduced to the mean shift and mode 2 to the spectral temperature. The mean shift mode 1 is more significant (order $v_0$ term) and arises from the Doppler shift introduced by the net CoM velocity. A bulk fluid flow gives a net CoM velocity, experimental measurements of bulk flows using the mode 1 shifts in the neutron spectral mean have been performed at OMEGA\cite{Mannion2018} and NIF\cite{Hatarik2015}. However, bulk fluid flows are not the only mechanism to introduce drifts to the reacting ions. For example, thermal gradients set up ion drifts which introduce a spectral shift \cite{Appelbe2017}. The less significant mode 2 term is due to a cross term between the energy shift and kinematic beaming of the Doppler shift. While the emission is isotropic in the CoM frame, a net CoM velocity will cause beaming in its direction in the lab frame - see \cref{eqn:anisoreactivity}. Note that if we work in the rest frame of the reacting ions then these terms will vanish. Finally, a mode 2 anisotropy in the reacting ions' distribution of CoM velocities will induce a mode 2 spectral anisotropy. This is given through the $\langle  \mathbf{M}(\vec{v}_{cm}) \rangle$ term. The axes of the spectral anisotropy will be aligned with those of the CoM velocity distribution.
	
	These expansions share great similarity with the hydrodynamic theory of spectral moments by Munro \cite{Munro2016}. A hydrodynamic plasma is one in which each fluid element has a Maxwellian distribution with every ion species having the same temperature and fluid velocity. For a hydrodynamic plasma, the distribution of $v_{cm}$ is determined by both temperature and fluid velocity while the distribution of $K$ is a function of temperature only. We see that the lowest order terms in the spectral moments only exhibit spherical harmonic modes up to 2. Indeed, a macroscopic hydrodynamic plasma can exhibit the same angular dependence as a anisotropic kinetic plasma. In the following section we will show how we can differentiate between ensembles of hydrodynamic and some classes of kinetic plasma. 
		
	\subsection{Macroscopic scale}
	Within a fusing plasma, fusion spectra are built up from the products from reactions occurring in different regions of velocity and coordinate space. Each local point emits a spectrum depending on averages over velocity space. The fusing plasma as a whole emits a spectrum which is the culmination of many different, local spectra, c.f. \cref{fig:FrameDiagram}. Thus, the statistics of this averaging is of great importance to the spectral interpretation.

	We have calculated the moments from a locally uniform region of the fusing plasma or ``fluid element''. However, the fluid elements which make up the fusing plasma may have non-zero velocity themselves, if the mean particle velocity of the distribution is non-zero. Boosting the product spectra back to the lab frame introduces a Doppler shift. Literature results for Maxwellian distributions with non-zero fluid velocity \cite{Appelbe2011,Munro2016,Murphy2014} have shown that these Doppler shifts introduce an anisotropic shift in the mean and additional broadening to the fusion peak. To first order in fluid velocity (assuming $u \ll v_0$), the non-relativistic modification to the spectral moments are given by:
	\begin{align}
		\Delta E_m &= \{\Delta E\} + m_3 v_0 \langle u_\parallel \rangle + ... \\
		T_{s,m} &= \{ T_s \} + (m_1+m_2)\mbox{Var}( u_\parallel) + ...
	\end{align}
	where $\Delta E_m$ and $T_{s,m}$ are the measured spectral moments, $\Delta E$ and $T_s$ are the fluid rest frame spectral moments, $u_\parallel$ is the component of the fluid velocity along the emission direction. As the action of the fluid velocity is the same for any distribution, these results are general -- for reactions of differing species the fluid velocity is the mass weighted drift velocity. The anisotropic components of the fluid velocity mean shift \cite{Hatarik2018,Mannion2018} and the fluid velocity variance\cite{Woo2018} can be extracted given sufficient measurements. We can therefore compare the isotropic terms in the ($\Delta E$, $T_s$) space. These extracted isotropic moments are not equal to those calculated without fluid velocity as there will remain an isotropic component to the fluid velocity variance which cannot be removed, this will inflate any measured spectral temperature.
	
	\begin{figure}[htp]
		\centering
		\includegraphics*[width=\columnwidth]{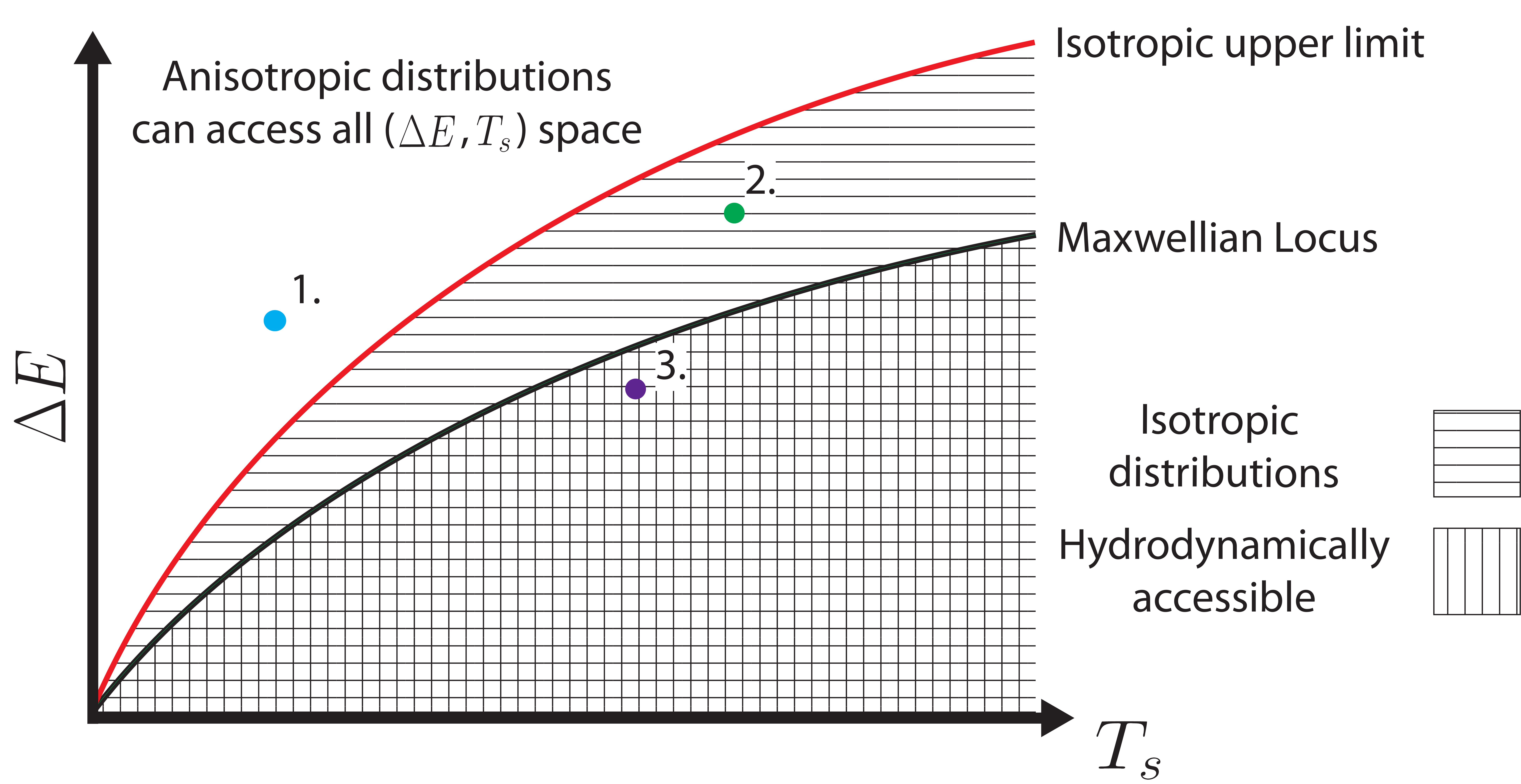}
		\caption{A diagram showing the ($\Delta E$, $T_s$) space and the various constraints on the form of the reactant distributions. For example, point 1 can only have been produced by anisotropic velocity distributions. Point 2 must have been produced by either isotropic and anisotropic distributions - the isotropic distributions must include a majority of momenta-matched reactions due to the proximity to the isotropic upper limit. Finally, point 3 is the only point which could have been produced by a collection of Maxwellians. While the above schematic is general, note that the exact positions of the constraint curves in ($\Delta E$, $T_s$) space depend on the specific reaction. }
		\label{fig:SpecMomLimiting}
	\end{figure}
	
	If, in the fluid rest frame, the energy shift is an increasing ($\Delta E' > 0$), concave ($\Delta E'' < 0$) function of spectral temperature, the effects of burn-averaging and fluid velocity variance lead to the following inequalities:
	\begin{align}
		\Delta E_m^{\mathrm{iso}} &\approx \langle \Delta E \rangle \leq \Delta E(T_{s,m}) \ , \\
		T^{\mathrm{min}}_{s,m} &\geq \langle T_s \rangle \ ,
	\end{align}
	Therefore, these effects move the observed spectral moments below the rest frame moments curve. The single temperature Maxwellian locus is a increasing, concave function and therefore an ensemble of hydrodynamic plasma will have spectral moments below the Maxwellian locus. Observations of spectral mean shifts and temperatures above the Maxwellian locus are therefore direct evidence of non-equilibrium distributions. As discussed above, spectral moments which lie above the isotropic momenta-matched limit must have originated from anisotropic reactant distributions. Below the Maxwellian locus, one must consider whether the deviation can be reasonably explained by the inclusion of fluid velocity and temperature variance. These variances can be quantified with hydrodynamics simulations. Large discrepancies can also be an indication of deviation from the equilibrium reactant distributions. The various classifying regions of the ($\Delta E$, $T_s$) space are summarised in \cref{fig:SpecMomLimiting}.

	In this work, we have been describing the fusion product birth spectra. The primary spectrum that reaches a detector can deviate from the birth spectrum due to transport effects such as scattering and attenuation. Munro\cite{Munro2016} provides formulae for how these transport effects can be quantified. If unaccounted for, these transport effects modify the spectral moments of the primary spectra and the constraints on the birth spectra ($\Delta E$, $T_s$) can be violated. These transport effects are depend on the amount of scattering and therefore are more important at high areal density.
	
	\section{Conclusions}
	
	The fusion product spectrum is sensitive to the reactant velocity distribution function. Ion kinetic effects can distort these distributions away from the equilibrium Maxwellian. Therefore, fusion product spectroscopy can be used to detect the presence of ion kinetic effects. This was achieved by studying the behaviour of the spectral moments: the spectral mean shift and energy variance (or spectral temperature). In this work, an analysis based on the work of Ballabio\cite{Ballabio1998} was used to investigate the transformation from reactant distribution functions to the spectral moment space. This allows one to relate features of the reactant distributions to the resultant spectral moments, these are intimately linked through reaction kinematics.
	
	Monoenergetic distributions were used to evaluate the limiting cases for isotropic distributions. When the reactant velocities were momenta matched, the mean shift was maximised for a given spectral temperature. When one of the reactants was stationary, the mean shift was minimised for a given spectral temperature. Since all isotropic distributions can be decomposed into a series of monoenergetic shells, we showed that these cases provide limits for any isotropic reactant distribution. The Maxwellian locus lies between these limits.
	
	General expressions for the directionally dependent spectral moments for anisotropic distributions were derived. These showed that the lowest order terms contain spherical harmonics modes up to 2. The spectral anisotropy depends on the distribution of centre of mass velocities as these introduce Doppler shifts in the fusion products. The 4-$\pi$ averaged moments exhibit the same form as that of isotropic distributions but anisotropic distributions do not exhibit constraints in the ($\Delta E, T_s$) space. Future work will explore the relationships between the form of the anisotropic distributions and their moments.
	
	At the macroscopic level, we considered the effect of a distribution of fluid elements with varying isotropic reactant distributions and fluid velocities. As the fluid velocities simply provide an additional Doppler effect, the upper, momenta-matched limit still applied even in this case. This analysis also showed that the Maxwellian locus defines an upper limit for any hydrodynamic ensemble of fluid elements. Indeed, recent Inertial Confinement Fusion experiments have shown primary fusion spectral moments which are incompatible with a Maxwellian velocity distribution description as the results lie above the Maxwellian locus \cite{Mannion_inprep2}.
	
	\section*{Acknowledgements}
	Sandia National Laboratories is a multimission laboratory managed and operated by National Technology \& Engineering Solutions of Sandia, LLC, a wholly owned subsidiary of Honeywell International Inc., for the U.S. Department of Energy's National Nuclear Security Administration under contract DE-NA0003525.  This paper describes objective technical results and analysis.  Any subjective views or opinions that might be expressed in the paper do not necessarily represent the views of the U.S. Department of Energy or the United States Government.
	
	\appendix
	
	\section{Monoenergetic Distributions}\label{appendix:monodist}
	
	In this appendix, we will consider isotropic monoenergetic distributions and derive formulae for their spectral moments. These distributions are shells in velocity space and therefore any general isotropic distribution is a superposition of these shells. We consider reactions between two shells, each defined by their particle speeds, $v_1$ and $v_2$. Since the cross section is a function of relative velocity only, we will also change to a set of variables which better describe the range of possible relative velocities. The maximum ($v_1+v_2$) and minimum ($v_1-v_2$) relative velocities are used to define the following:
	\begin{align}
	K_{\mathrm{max}} &= \frac{1}{2}m_{12}(v_1+v_2)^2 \ , \\
	\chi &= \frac{v_1-v_2}{v_1+v_2} \ ,
	\end{align}
	where we note that $\chi$ is a signed quantity depending on the ordering of $v_1$ and $v_2$ and takes values between -1 and +1. The benefit of this variable set is that one can maintain a fixed $K_{\mathrm{max}}$, which sets the cross section for the highest energy reaction, and vary $\chi$, which controls the ``character" of the reactions i.e. the partition of kinetic energy between the reactant species. The average relative kinetic energy, spectral temperature and mean shift are given by the following expressions:
	\begin{subequations}
		\begin{align}
		\langle K \rangle &= K_{\mathrm{max}} \mathcal{L}_K \ , \\
		T^s_{12} &= K_{\mathrm{max}} \mathcal{L}_T \ , \\
		\Delta E &= K_{\mathrm{max}} (\alpha_K \mathcal{L}_K + \alpha_T \mathcal{L}_T) \ ,
		\end{align}
		where we have defined scaling functions:
		\begin{align}
		\mathcal{L}_K &\equiv \frac{\int^{1}_{|\chi|} x^4 \sigma(x^2 K_{\mathrm{max}}) dx}{\int^{1}_{|\chi|} x^2 \sigma(x^2 K_{\mathrm{max}}) dx} \ , \\
		\mathcal{L}_T &\equiv \frac{2}{3}\left[\frac{1}{4}\frac{m_1+m_2}{m_{12}}(1+\chi^2)+\frac{1}{2}\frac{m_1-m_2}{m_{12}}\chi-\mathcal{L}_K\right] \ ,
		\end{align}
		and, in the $\mathcal{L}_K$ integrals, $x = \sqrt{K/K_{\mathrm{max}}}$.
	\end{subequations}
	We have also defined a new moment coefficient,
	\begin{equation}
	\alpha_T = \frac{3}{m_1+m_2}\alpha_V \ ,
	\end{equation}
	such that $T_s$ can be used over $\langle v_{cm}^2 \rangle$ by including the appropriate coefficient. The function $\mathcal{L}_K$ gives the average relative kinetic energy, $\langle K \rangle$, as a fraction of the maximum relative kinetic energy, $K_{max}$. The function $\mathcal{L}_T$ gives $2/3$ times the average centre of mass energy as a fraction of the maximum relative kinetic energy, $K_{max}$, where the factor of $2/3$ comes from equipartition ($T = \frac{2}{3}E$).
	
	A few key observations can be made from these general expressions:
	
	Firstly, the Coulomb barrier penetrability exponentially suppresses the contributions to $\mathcal{L}_K$ at low values of $K$ (or $x$). For low $K_{\mathrm{max}}$, this will lead to $\mathcal{L}_K \rightarrow 1$. In velocity space, only antipodal points (head-on collisions) lead to successful fusion reactions. In contrast, particles which are in the vicinity of each other in velocity space have negligible probability of reaction. As $K_{\mathrm{max}}$ increases, reactions with $K < K_{\mathrm{max}}$ have sufficient energy to overcome the Coulomb barrier leading to a decrease in $\mathcal{L}_K$. 
	
	Secondly, for a given $K_{\mathrm{max}}$, the minimum of $\mathcal{L}_T$ occurs at (or just above) the momenta matched condition for the species velocities:
	\begin{equation}
	\chi_{\mathrm{min}} = \frac{m_2-m_1}{m_1+m_2}+\frac{2m_{12}}{m_1+m_2}\left.\frac{\partial \mathcal{L}_K}{\partial \chi}\right|_{\chi = \chi_{\mathrm{min}}} \ .
	\end{equation}
	As discussed in the previous point, the Coulomb barrier penetrability suppresses contributions from low $x$ and thus $\frac{\partial \mathcal{L}_K}{\partial \chi} \ll 1$ for small $\chi$. We arrive at the physically intuitive result that when the species momenta are matched, we obtain the lowest Doppler broadening as many fusion reactions occur with low or vanishing CoM velocity. In this case, the spectral temperature and mean shift are given by:
	\begin{align}
	T^s_{12} &= \frac{2}{3}K_{\mathrm{max}} (1-\left.\mathcal{L}_K\right|_{\chi = \chi^*}) \ , \\
	\Delta E &= \frac{2}{3}K_{\mathrm{max}} \left[\left(\frac{3}{2}\alpha_K -  \alpha_T\right) \left.\mathcal{L}_K\right|_{\chi = \chi^*} + \alpha_T\right] \ , \\
	\chi^* &\equiv \frac{m_2-m_1}{m_1+m_2} \ .
	\end{align}
	Plots and functional fits to $\left.\mathcal{L}_K\right|_{\chi = \chi^*}$ are given below.
	
	Finally, large differences in the velocities of reactants ($\chi \rightarrow \pm 1$) reduces the range of relative velocities possible, leading to reactions only occurring at $K_{\mathrm{max}}$ i.e. $\mathcal{L}_K \rightarrow 1$. In this beam-target limit, $\mathcal{L}_K$ and $\mathcal{L}_T$ both tend to constants and therefore the spectral temperature and mean shift become directly proportional. The constants of proportionality are as follows:
	\begin{align}
	\Delta E &= k T^s_{12}\ , \\
	k &= \begin{cases}
	\frac{3m_2}{2m_1}\alpha_K + \alpha_T \ , \ \chi = +1 \ , \\
	\frac{3m_1}{2m_2}\alpha_K + \alpha_T \ , \ \chi = -1 \ ,
	\end{cases}
	\end{align}
	
\begin{figure}[htp]
		\centering
		\includegraphics*[width=0.99\columnwidth]{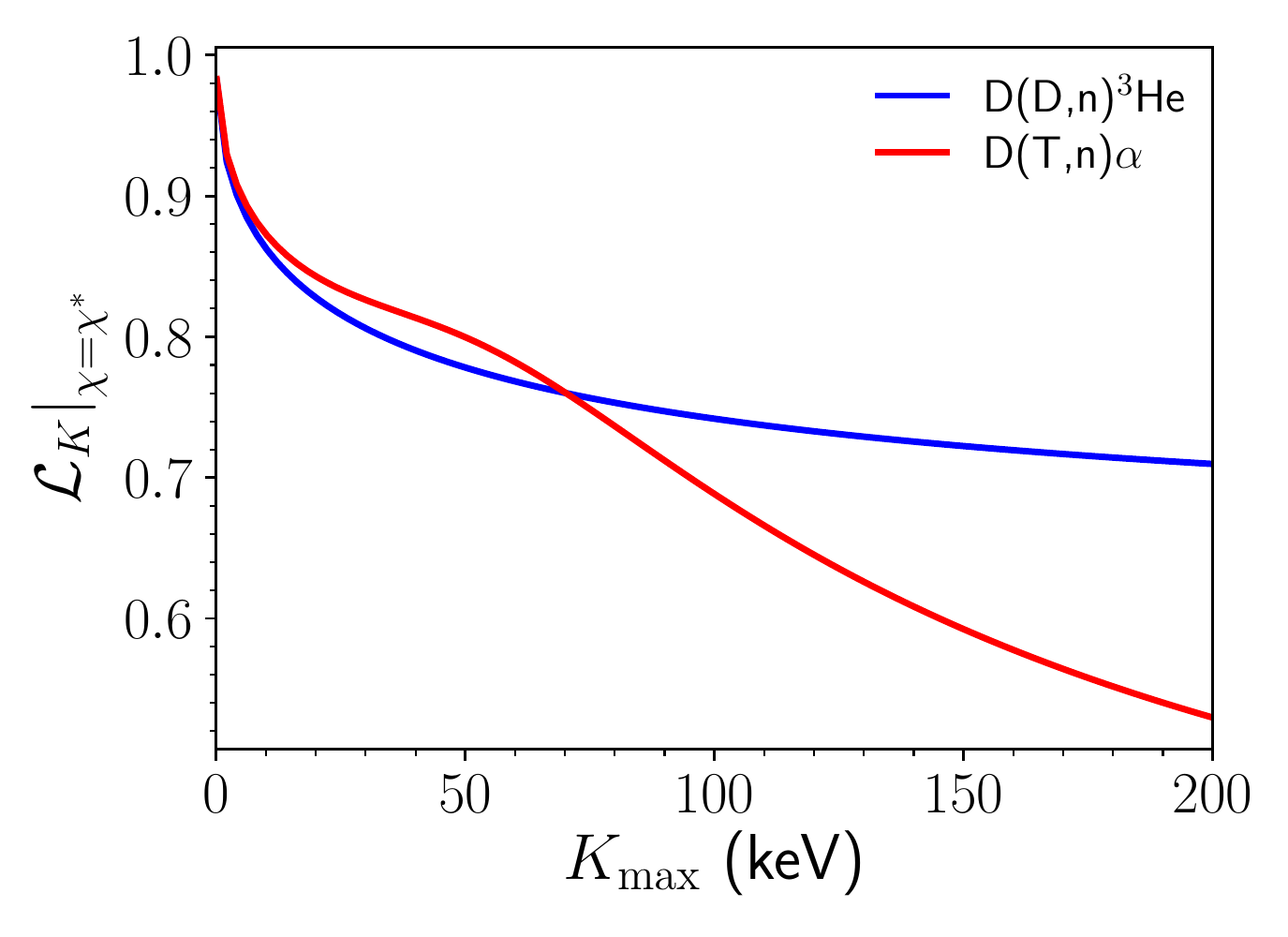}
		\caption{The momenta-matched relative kinetic energy scaling as a function of $K_{\mathrm{max}}$ for D(D,n)$^3$He and D(T,n)$\alpha$. In can be seen that the resonance in the DT cross section at $\sim$ 60 keV has a strong effect on the relative kinetic energy.}
		\label{fig:LK_mommatch}
	\end{figure}

	The Bosch-Hale\cite{Bosch1992} fusion cross sections of D(D,n)$^3$He and D(T,n)$\alpha$ were used to evaluate the momenta-matched relative kinetic energy scaling as a function of $K_{\mathrm{max}}$. The results of this evaluation are shown in \cref{fig:LK_mommatch}. 
	
	A rational function fit of the following form:
	\begin{equation}\label{eqn:rationalfuncfit}
	\mathcal{L}_K(K_{\mathrm{max}},\chi = \chi^*) = \frac{1+\sum_{i=1}^{4}p_i K_{\mathrm{max}}^i}{1+\sum_{j=1}^{4}q_j K_{\mathrm{max}}^j}
	\end{equation}
	is provided for D(D,n)$^3$He and D(T,n)$\alpha$. $K_{\mathrm{max}}$ is in units of keV and the fit was performed using the data shown in \cref{fig:LK_mommatch} in the range 0 to 200 keV. The fitting parameters, $q_i$ and $p_i$, are given in \cref{table:LKfit}.
	
	\begin{table}[ht]\caption{Coefficients for rational fit given in \cref{eqn:rationalfuncfit}. This fit has a maximal absolute error of 0.003 within range 0 to 200 keV.}
		\begin{center}
			\begin{tabular}{c | c | c}
				& D(D,n)$^3$He & D(T,n)$\alpha$ \\
				\hline
				$p_1$ & 6.79e+0 & 3.18e+0 \\
				$p_2$ & 1.66e+0 & 2.70e-2 \\
				$p_3$ & 4.35e-2 & 6.02e-4 \\
				$p_4$ & 1.49e-4 & 5.06e-6 \\
				\hline
				$q_1$ & 7.08e+0 & 3.40e+0 \\
				$q_2$ & 1.91e+0 & 6.33e-2 \\
				$q_3$ & 5.89e-2 & -6.79e-5 \\
				$q_4$ & 2.33e-4 & 1.56e-5
			\end{tabular}
		\end{center}
		\label{table:LKfit}
	\end{table}
	
	When calculating the minimum of $\mathcal{L}_T$, we assumed that $\frac{\partial \mathcal{L}_K}{\partial \chi}$ was small due to Coulomb barrier suppression of low energy reactions. Numerical calculation of the derivative, within the [0,200] keV range, supports this argument.
	
	It is noted for D(D,n)$^3$He and D(T,n)$\alpha$ that, for any given $\chi$, $\Delta E$ is an increasing ($\Delta E' > 0$), concave ($\Delta E'' < 0$) function of $T_s$. It also follows that if $\langle K \rangle$ is an increasing, concave function of $T_s$ then $\Delta E$ will also be. This can be related to requirements on $\mathcal{L}_K$ and the fusion cross section. In order for $\langle K \rangle(T_s)$ to be an increasing function:
	\begin{equation}
	0 < \frac{\partial }{\partial K_{\mathrm{max}}}\left[K_{\mathrm{max}}\mathcal{L}_K\right] < 1 \ ,
	\end{equation}
	and in order for $\langle K \rangle(T_s)$ to also be a concave function:
	\begin{equation}
	\frac{\partial^2 }{\partial K_{\mathrm{max}}^2}\left[K_{\mathrm{max}}\mathcal{L}_K\right] < 0 \ .
	\end{equation}
	Since $\mathcal{L}_K$ is defined as integrals over the fusion cross section, these inequalities are constraints on the shape of the cross section.

	\section{Expressions for anisotropic moments}\label{appendix:aniso}
	
	It is expected that anisotropic velocity distributions will give rise to anisotropic spectra. Evaluating spectral moments where we consider the anisotropy of emission requires a more general expression for the reaction rate weighting which is emission direction dependent. First, we must start with the full differential reaction rate and the directional neutron energy moments:
	
	\begin{align}\label{e:1.1}
	dR &=  v_{r}\frac{d\sigma}{d\Omega}F\left(\vec{v}_{cm},\vec{v}_{r}\right)d^{3}\vec{v}_{cm}d^{3}\vec{v}_{r}d\Omega_{s} \ , \\
	\langle E_3^n \rangle &= \frac{\int E_3^n \delta(\hat{v}_3 - \Omega) dR}{\int \delta(\hat{v}_3 - \Omega) dR} \ ,
	\end{align}
	where
	\begin{itemize}
		\item $\frac{d\sigma}{d\Omega}\left(v_{r},\hat{v}_{r}\cdot\hat{u}_{3}\right)$ - differential cross-section for the reaction
		\item $F\left(\vec{v_{cm}},\vec{v_{r}}\right)$ - joint probability distribution of $\vec{v}_{cm}$ and $\vec{v}_{r}$ given by the product of the reactant velocity distributions, $f_1(\vec{v}_1[\vec{v}_{cm},\vec{v}_{r}])f_2(\vec{v}_2[\vec{v}_{cm},\vec{v}_{r}])$.
		\item $\vec{u}_{3}$ - fusion product velocity in CoM frame
	\end{itemize}
	
	Where the Dirac delta ensures that we only consider emission along a particular direction, $\Omega$. Transformation of variables is required to integrate out this Dirac delta and produce a tractable form for theoretical analysis. The following steps outline the required procedure:
	\begin{enumerate}
		\item Assume $d\Omega_{s} = d\Omega_{u}$ where $d\Omega_{u}$ is the solid angle of the vector $\vec{u}_{3}$
		\item Transform $\vec{v}_{r}$ to spherical coordinates
		\begin{equation*}
		d^{3}\vec{v}_{r} = v_{r}^{2}dv_{r}d\Omega_{r}
		\end{equation*}
		\item Transform $v_{r}$ to $u_{3}$ using
		\begin{eqnarray*}
		u_{3} &=& \sqrt{\eta\left(Q+K\right)}\\
		\eta &=& \frac{2m_{4}}{m_{3}\left(m_{3}+m_{4}\right)}\\
		K &=& \frac{1}{2}m_{12}v_{r}^{2}\\
		dv_{r} &=& \frac{2}{m_{12}\eta}\frac{u_{3}}{v_{r}}du_{3}
		\end{eqnarray*}
		\item Transform $du_{3}d\Omega_{u}$ to Cartesian co-ordinates
		\begin{equation*}\label{e:1.4}
		du_{3}d\Omega_{u} = \frac{1}{u_{3}^{2}}d^{3}\vec{u}_{3}\nonumber
		\end{equation*}
		\item Transform $\vec{u}_{3}$ to $\vec{v}_{3}$ using
		\begin{eqnarray*}
		\vec{u}_{3}&=& \vec{v}_{3}-\vec{v}_{cm}\\
		d^{3}\vec{u}_{3} &=& d^{3}\vec{v}_{3}
		\end{eqnarray*}
		\item Transform $\vec{v}_{3}$ to spherical co-ordinates and transform $v_{3}$ to $E_{3}$
		\begin{eqnarray*}
		E_{3}&=& \frac{1}{2}m_{3}v_{3}^{2}\\
		d^{3}\vec{v}_{3} &=& \sqrt{\frac{2E_{3}}{m_{3}^{3}}}dE_{3}d\hat{v}_{3} 
		\end{eqnarray*}
		\item Transform $E_{3}$ to $u_{3}$ using
		\begin{eqnarray*}
		\sqrt{E_{3} }&=& \sqrt{\frac{m_{3}}{2}}v_{cm}\left[\hat{v}_{cm}\cdot\hat{v}_{3}\pm\sqrt{\frac{u_{3}^{2}}{v_{cm}^{2}}-1+\left(\hat{v}_{cm}\cdot\hat{v}_{3}\right)^{2}}\right] 
		\end{eqnarray*}
		allowing us to compute the Jacobian
		\begin{eqnarray*}
		dE_{3} &=& \sqrt{2m_{3}E_{3}}\frac{u_{3}}{v_{cm}}\frac{1}{\sqrt{\frac{u_{3}^{2}}{v_{cm}^{2}}-1+\left(\hat{v}_{cm}\cdot\hat{v}_{3}\right)^{2}}}du_{3}
		\end{eqnarray*}
		\item Finally, transform $u_{3}$ to $v_{r}$ and then transform $dv_{r}d\Omega_{r}$ from spherical to cartesian co-ordinates
		\begin{equation*}
		du_{3}d\Omega_{r} = \frac{m_{12}\eta}{2}\frac{1}{u_{3}v_{r}}d^{3}\vec{v}_{r} \nonumber
		\end{equation*}
	\end{enumerate}
	\begin{widetext}
	Thus, we convert the differential reaction rate to the following form:
	\begin{equation}
	dR  = \frac{\left[u_{3}^{2}-v_{cm}^{2}+2\left(\vec{v}_{cm}\cdot\hat{v}_3\right)^{2}+2\left(\vec{v}_{cm}\cdot\hat{v}_3\right)\sqrt{u_{3}^{2}-v_{cm}^2+\left(\vec{v}_{cm}\cdot\hat{v}_3\right)^{2}}\right]}{u_3\sqrt{u_{3}^{2}-v_{cm}^2+\left(\vec{v}_{cm}\cdot\hat{v}_3\right)^{2}}} \times v_r\frac{d\sigma}{d\Omega}F\left(\vec{v_{cm}},\vec{v_{r}}\right)d^{3}\vec{v_{cm}}d^{3}\vec{v_{r}}d\hat{v}_3 \label{e:1.10}
	\end{equation}
	and in this form the requirement that the emission direction is along $\Omega$ can be trivially satisfied by substituting $\hat{v}_3$ for $\Omega$. We will use this expression to build a general theory for anisotropic spectral moments. In particular, we are interested in the term introduced by specifying a particular emission direction for the n'th order energy moment:
	\begin{equation}
	A_n(K,v_{cm},\Omega) = \frac{\left[u_{3}^{2}-v_{cm}^{2}+2\left(\vec{v}_{cm}\cdot\Omega\right)^{2}+2\left(\vec{v}_{cm}\cdot\Omega\right)\sqrt{u_{3}^{2}-v_{cm}^2+\left(\vec{v}_{cm}\cdot\Omega\right)^{2}}\right]^{n+1}}{u_3\sqrt{u_{3}^{2}-v_{cm}^2+\left(\vec{v}_{cm}\cdot\Omega\right)^{2}}}  \label{e:1.11}
	\end{equation}
	This will be used in the following sections to calculate the isotropic and anisotropic spectral moments. 
	\end{widetext}
	
	\subsection{$4\pi$-averaged moments}
	First, we will show that \Cref{e:1.10,e:1.11} return the familiar isotropic moment results. To do this we will assume an isotropic differential cross section and integrate over all emission directions, $\Omega$. Separating out the angularly dependent terms and assuming that $\hat{v}_{cm}$ is in the polar direction without loss of generality, the integration over $d\Omega$ involves:
	\begin{eqnarray}
	&&\frac{1}{4\pi}\int A_n(K,v_{cm},\Omega)d\Omega\nonumber\\ 
	&&= v_{cm}^{2n} \cdot \frac{1}{2} \int\limits_{-1}^{1}\frac{\left[a+2x^{2}+2x\sqrt{a+x^{2}}\right]^{n+1}}{\sqrt{a+1}\sqrt{a+x^{2}}}dx\nonumber
	\end{eqnarray}
	where $a = \frac{u_{3}^{2}}{v_{cm}^{2}}-1 \geq 0$ and $x = \hat{v}_{cm} \cdot \Omega$. The above integral can be performed analytically for any given energy moment order, $n$. For n = 0, 1 and 2, the integral over $dx$ has the following values
	\begin{align}
	n &= 0: && 1\nonumber\\
	n &= 1: && u_3^2+v_{cm}^2\nonumber\\
	n &= 2: &&\frac{1}{3}\left(3u_{3}^{4}+10u_{3}^{2}v_{cm}^{2}+3v_{cm}^{4}\right) \nonumber
	\end{align}
	The results for the isotropic moments for $n$ = 0, 1 and 2 are then:
	\begin{align}
	n &= 0: && \int v_{r}\sigma F\left(\vec{v_{cm}},\vec{v_{r}}\right)d^{3}\vec{v_{cm}}d^{3}\vec{v_{r}} = \langle \sigma v \rangle \nonumber\\
	n &= 1: && \frac{m_3}{2}\langle u_3^2+v_{cm}^2 \rangle = E_0 + \frac{m_4}{m_3+m_4} \langle K \rangle + \frac{1}{2}m_3 \langle v_{cm}^2 \rangle \nonumber\\
	n &= 2: && \frac{m_3^2}{12}\langle 3u_{3}^{4}+10u_{3}^{2}v_{cm}^{2}+3v_{cm}^{4}\rangle = E_0^2\nonumber \\
	& && + E_0\left(\frac{2m_4}{m_3+m_4}\langle K \rangle + \frac{5}{3}m_3 \langle v_{cm}^2 \rangle\right) \nonumber \\
	& && + \left(\frac{m_4}{m_3+m_4}\right)^2\langle K^2 \rangle + \frac{5 m_3m_4}{3(m_3+m_4)} \langle v_{cm}^2 K \rangle \nonumber \\
	& && + \frac{m_3^2}{4}\langle v_{cm}^4 \rangle \nonumber
	\end{align}
	where $v_0 = \sqrt{\eta Q}$ and $E_0 = m_3v_0^2/2$ are the velocity and energy of the fusion product for $K = v_{cm} = 0$. From these moments we can get the first two cumulants of the spectrum:
	\begin{align}
		\langle E_3 \rangle &= E_0 + \frac{m_4}{m_3+m_4} \langle K \rangle + \frac{1}{2}m_3 \langle v_{cm}^2 \rangle \ , \\
		\langle E_3^2 \rangle - \langle E_3 \rangle^2 &= \frac{2}{3}m_3E_0 \langle v_{cm}^2 \rangle \\
		&+ \frac{m_4}{m_3+m_4}\left(\langle K^2 \rangle - \langle K \rangle^2\right) \nonumber \\
		&+ \frac{1}{2}m_3\left(\langle v_{cm}^4 \rangle - \langle v_{cm}^2 \rangle^2\right) \nonumber \\
		&+ \frac{m_3m_4}{m_3+m_4}\left(\frac{5}{3}\langle v_{cm}^2 K \rangle-\langle v_{cm}^2 \rangle \langle K \rangle\right) \ . \nonumber
	\end{align}
	These match the classical kinematics results of Ballabio\cite{Ballabio1998} and Brysk\cite{Brysk1973}. We have shown that the form of the 4-$\pi$ averaged moments for anisotropic distributions match those of purely isotropic distributions. Therefore, if it is possible to infer the 4-$\pi$ averaged moments, we can know we are comparing the same $v_{cm}^2$ and $K$ moments of the reactant velocity distribution regardless of any anisotropy. 
	
	\subsection{Anisotropic moments}
	In experiments, detectors measure spectra along particular lines of sight so we must understand the anisotropy that anisotropic distributions introduce to these measurements. With sufficient measurements, modes of this anisotropy can be isolated and analysed individually. This analysis has already been performed for the effect of fluid velocity which introduces a $L=1$ mode \cite{Munro2016,Hatarik2018,Mannion2018}.
	
	For the anisotropic moments, no integration over emission direction is performed. Instead, $\Omega$ is treated as known constant unit vector and integration over the velocity space of $\vec{v}_{cm}$ and $\vec{v}_r$ is performed. To simplify the problem, we consider a small expansion parameter, $1/v_0$, for expansions of $A_n$. We used Mathematica to accurately perform the expansions to the required order in $v_0$. We will denote directionally dependent moments as $\{x\}$.
	First, we must determine the directionally dependent reactivity as it is used to normalise all energy moments. This involves expansion of the function $A_0$:
	\begin{align}
		\{\sigma v \} &= \langle \sigma v \rangle \langle A_0 \rangle \\
		&= \langle \sigma v \rangle\left(1+\frac{2}{v_0} \langle \vec{v}_{cm} \cdot \Omega \rangle \right.\\
		&\left.+ \frac{3}{2v_0^2} \langle (\vec{v}_{cm} \cdot \Omega)^2 - \frac{1}{3}v_{cm}^2 \rangle + \mathcal{O}\left(\frac{1}{v_0^3}\right)\right)  \ . \nonumber
	\end{align}
	As $\Omega$ is a constant vector, it can be taken outside the reaction rate average. This allows us to use a symmetric traceless tensorial notation to easily separate terms by their angular mode:
	\begin{align}
	\{\sigma v \} &= \langle \sigma v \rangle\left(1+\frac{2}{v_0} \Omega^T \langle \vec{v}_{cm} \rangle \right. \\
	&\left. + \frac{3}{2v_0^2} \Omega^T \langle \mathbf{M}(\vec{v}_{cm}) \rangle \Omega + \mathcal{O}\left(\frac{1}{v_0^3}\right)\right) \nonumber \ ,
	\end{align}
	where $M$ is the symmetric, traceless rank 2 tensor given by:
	\begin{align}
	\mathbf{M}(\vec{x})  &= \vec{x} \otimes \vec{x} - \frac{|\vec{x}|^2}{3} \mathds{1} \ ,
	\end{align}
	where $\otimes$ denotes the outer product and $\mathds{1}$ the identity matrix. The spherical harmonic mode of each contribution is given by the rank of the tensor or, in an alternative view, the order of term in $\Omega$. As expected all anisotropy is introduced by the distribution of CoM velocities. This is because $K$ has an effect on the outgoing energy of the product but not its direction (for an isotropic differential cross section). All further moments follow a similar procedure but are normalised by the reactivity. For the directionally dependent mean energy:
	\begin{equation}
		\{E_3\} = \frac{\langle A_1 \rangle}{\langle A_0 \rangle} \ ,
	\end{equation}
	Again, expanding in small parameter $1/v_0$ allows us to find the directionally dependent energy shift:
	\begin{align}
		\{ \Delta E \} &= \{ E_3 \} - E_0 \\
		&= \frac{m_4}{m_3+m_4} \langle K \rangle + \frac{1}{2}m_3 \langle v_{cm}^2 \rangle  \\
		&+ m_3 v_0 \hat{\Omega}^T \langle \vec{v}_{cm} \rangle \nonumber \\
		&- m_3 \hat{\Omega}^T \mathbf{M}(\langle \vec{v}_{cm} \rangle) \hat{\Omega} - \frac{1}{3}m_3|\langle \vec{v}_{cm} \rangle|^2 \nonumber \\
		&+ 3 m_3 \hat{\Omega}^T \langle \mathbf{M}(\vec{v}_{cm}) \rangle \hat{\Omega} + \mathcal{O}\left(\frac{1}{v_0}\right)  \ . \nonumber
	\end{align}
	Finally, the directionally dependent spectral temperature is given by:
	\begin{align}
	\{T_s\} &= \frac{m_1+m_2}{m_3^2v_0^2}\left(\{ E_3^2 \} - \{ E_3 \}^2\right) \\
	&= \frac{m_1+m_2}{3} \langle v_{cm}^2 \rangle \\
	&+ (m_1+m_2)\hat{\Omega}^T \langle  \mathbf{M}(\vec{v}_{cm}) \rangle \hat{\Omega} \nonumber \\
	&- (m_1+m_2) \hat{\Omega}^T \mathbf{M}(\langle \vec{v}_{cm} \rangle) \hat{\Omega} - \frac{m_1+m_2}{3}|\langle \vec{v}_{cm} \rangle|^2 \nonumber \\
	&+ \mathcal{O}\left(\frac{1}{v_0}\right)  \ . \nonumber
	\end{align}
	To better illustrate the form of the mode 2 terms consider the eigenvalues, $\lambda_{cm}^{(i)}$, of $\mathbf{M}(\vec{v}_{cm})$ which sum to 0 by definition. An isotropic distribution will have no preferred direction and therefore all eigenvalues must be equal to 0. If these eigenvalues are not all equal then the mean shift and spectral temperature will have mode 2 anisotropy aligned with the eigenvectors, $\hat{w}_{cm}^{(i)}$, of $\mathbf{M}(\vec{v}_{cm})$. In a frame aligned with these eigenvectors:
	\begin{align}
	\hat{\Omega}^T \langle  \mathbf{M}(\vec{v}_{cm}) \rangle \hat{\Omega} &= \lambda_{cm}^{(1)} (\sin^2(\theta)\cos^2(\phi)-\cos^2(\theta)) \\
	&+ \lambda_{cm}^{(2)} (\sin^2(\theta)\sin^2(\phi)-\cos^2(\theta)) \ ,\nonumber
	\end{align}
	where we have used the traceless property of $M$ to write $\lambda^{(3)}_{cm} = -\lambda_{cm}^{(1)}-\lambda_{cm}^{(2)}$. The above expression is equivalent to the following spherical harmonic expansion:
	\begin{align}
	\hat{\Omega}^T \langle  \mathbf{M}(\vec{v}_{cm}) \rangle \hat{\Omega}&= \sqrt{\frac{2\pi}{15}}(\lambda_{cm}^{(1)}-\lambda_{cm}^{(2)})(Y_2^2+Y_2^{-2}) \\
	&-2\sqrt{\frac{\pi}{5}} (\lambda_{cm}^{(1)}+\lambda_{cm}^{(2)})Y_2^0 \ . \nonumber
	\end{align}
	Given the properties of the spherical harmonics, rotation of the axes will maintain an L=2 anisotropy.
	
	It is important to note that the 4-$\pi$ averaged moments are not equal to the isotropic part of the anisotropic moments. This is due to the reactivity normalisation. To illustrate this, compare the 4-$\pi$ average and isotropic mean energy:
	\begin{align}
		\langle E_3 \rangle_{4\pi} &=  \frac{\int d\Omega \langle A_1 \rangle}{\int d\Omega \langle A_0 \rangle} =  \int d\Omega \langle A_1 \rangle \ , \\
		\int \{E_3\} d\Omega &= \int d\Omega \frac{\langle A_1 \rangle}{\langle A_0 \rangle}  \ .
	\end{align}
	Therefore, one must account for the angular dependence of the reactivity/yield to extract the 4-$\pi$ averaged moment:
	\begin{align}
	\frac{\int \{\sigma v\}\{E_3\} d\Omega}{\langle \sigma v \rangle} &= \int d\Omega \langle A_1 \rangle = \langle E_3 \rangle_{4\pi} \ , \\
	\frac{\int \{Y_3\}\{E_3\} d\Omega}{Y_3} &= \int d\Omega \langle A_1 \rangle = \langle E_3 \rangle_{4\pi} \ .
	\end{align}
	
	\section*{References}
	\bibliography{MuCFRefs}
	
\end{document}